\documentstyle{mn}

\newif\ifAMStwofonts

\ifoldfss
  \ifCUPmtlplainloaded \else
    \NewTextAlphabet{textbfit} {cmbxti10} {}
    \NewTextAlphabet{textbfss} {cmssbx10} {}
    \NewMathAlphabet{mathbfit} {cmbxti10} {} 
    \NewMathAlphabet{mathbfss} {cmssbx10} {} 
  \fi
  \ifAMStwofonts
    \ifCUPmtlplainloaded \else
      \NewSymbolFont{upmath} {eurm10}
      \NewSymbolFont{AMSa} {msam10}
      \NewMathSymbol{\upi}     {0}{upmath}{19}
      \NewMathSymbol{\umu}     {0}{upmath}{16}
      \NewMathSymbol{\upartial}{0}{upmath}{40}
      \NewMathSymbol{\leqslant}{3}{AMSa}{36}
      \NewMathSymbol{\geqslant}{3}{AMSa}{3E}

       \let\le=\leqslant
       \let\ge=\geqslant
    \fi
  \fi
\fi 

\ifnfssone
  \newmathalphabet{\mathit}
  \addtoversion{normal}{\mathit}{cmr}{m}{it}
  \addtoversion{bold}{\mathit}{cmr}{bx}{it}
  \newmathalphabet{\mathbfit} 
  \addtoversion{normal}{\mathbfit}{cmr}{bx}{it}
  \addtoversion{bold}{\mathbfit}{cmr}{bx}{it}
  \newmathalphabet{\mathbfss} 
  \addtoversion{normal}{\mathbfss}{cmss}{bx}{n}
  \addtoversion{bold}{\mathbfss}{cmss}{bx}{n}
  \ifAMStwofonts
    \ifCUPmtlplainloaded \else
      \UseAMStwoboldmath
      \makeatletter
      \new@mathgroup\upmath@group
      \define@mathgroup\mv@normal\upmath@group{eur}{m}{n}
      \define@mathgroup\mv@bold\upmath@group{eur}{b}{n}
      \edef\UPM{\hexnumber\upmath@group}
      \new@mathgroup\amsa@group
      \define@mathgroup\mv@normal\amsa@group{msa}{m}{n}
      \define@mathgroup\mv@bold\amsa@group{msa}{m}{n}
      \edef\AMSa{\hexnumber\amsa@group}
      \makeatother
      \mathchardef\upi="0\UPM19
      \mathchardef\umu="0\UPM16
      \mathchardef\upartial="0\UPM40
      \mathchardef\leqslant="3\AMSa36
      \mathchardef\geqslant="3\AMSa3E

       \let\le=\leqslant
       \let\ge=\geqslant
    \fi
  \fi
\fi 

\ifnfsstwo
  \DeclareMathAlphabet{\mathbfit}{OT1}{cmr}{bx}{it}
  \SetMathAlphabet\mathbfit{bold}{OT1}{cmr}{bx}{it}
  \DeclareMathAlphabet{\mathbfss}{OT1}{cmss}{bx}{n}
  \SetMathAlphabet\mathbfss{bold}{OT1}{cmss}{bx}{n}
  \ifAMStwofonts
    \ifCUPmtlplainloaded \else
      \DeclareSymbolFont{UPM}{U}{eur}{m}{n}
      \SetSymbolFont{UPM}{bold}{U}{eur}{b}{n}
      \DeclareSymbolFont{AMSa}{U}{msa}{m}{n}
      \DeclareMathSymbol{\upi}{0}{UPM}{"19}
      \DeclareMathSymbol{\umu}{0}{UPM}{"16}
      \DeclareMathSymbol{\upartial}{0}{UPM}{"40}
      \DeclareMathSymbol{\leqslant}{3}{AMSa}{"36}
      \DeclareMathSymbol{\geqslant}{3}{AMSa}{"3E}

       \let\le=\leqslant
       \let\ge=\geqslant
    \fi
  \fi
\fi 

\ifCUPmtlplainloaded \else
  \ifAMStwofonts \else 
    \def\upi{\pi}
    \def\umu{\mu}
    \def\upartial{\partial}
  \fi
\fi

\title{On the second overtone stability among SMC Cepheids}
\author[G. Bono et al.]
{G. Bono,$^{1,2}$ F. Caputo,$^1$ and M. Marconi,$^3$\\  
$^1$ Osservatorio Astronomico di Roma, Via Frascati 33, 00040 Monte Porzio 
Catone, Italy\\
$^1$ Euroapen Southern Observatory, Casilla 19001, Santiago 19, Chile\\  
$^3$ Osservatorio Astronomico di Capodimonte, Via Moiariello 16, 80131 Napoli, 
Italy\\}   

\pagerange{\pageref{firstpage}--\pageref{lastpage}}
\pubyear{2000}

\begin{document}

\maketitle

\label{firstpage}

\begin{abstract}
We present a new set of Cepheid, full amplitude, nonlinear, convective 
models which are pulsationally unstable in the second overtone (SO). 
Hydrodynamical models were constructed by adopting a chemical composition 
typical for Cepheids in the Small Magellanic Cloud (SMC) and stellar 
masses ranging from 3.25 to 4 $M_\odot$.  
Predicted $\phi_{21}$ Fourier parameters agree, within current uncertainties, 
with empirical data for pure first and second overtone variables as well as 
for first/second overtone (FO/SO) double-mode Cepheids collected by Udalski 
et al. (1999a,b) in the SMC. On the other hand, predicted I band amplitudes 
are systematically larger than the observed ones in the short period range, 
but attain values that are closer to the empirical ones for 
$\log P_{SO} \ge -0.12$ and $\log P_{FO} \ge 0.1$. 
We also find, in agreement with empirical evidence, that the region 
within which both second and first overtones attain a stable limit cycle 
widens when moving toward lower luminosities. Moreover, predicted 
$P_{SO}/P_{FO}$ and $P_{FO}/P_F$ period ratios agree quite well with 
empirical period ratios for FO/SO and F/FO double-mode SMC Cepheids. 

Interestingly enough, current models support the evidence that pure SO 
Cepheids and SO components in FO/SO Cepheids are good distance indicators. 
In fact, we find that the fit of the predicted Period-Luminosity-Color
(V, V-I) relation to empirical SMC data supplies a distance modulus of 
$19.11\pm0.08$ mag. The same outcome applies to pure FO Cepheids and 
FO components in  FO/SO Cepheids, and indeed we find DM=$19.16\pm0.19$     
mag. Current distance estimates do not account for, within current 
uncertainties on photometry and reddening, the so-called short 
distance scale.    
\end{abstract}


\section{Introduction}

The existence of overtone pulsators among radial variables is a 
long-standing astrophysical problem (Eddington 1926; Ledoux \& 
Walraven 1958). On the basis of 
pulsating polytropic models, Schwarzschild (1941) suggested that 
among actual RR Lyrae variables are present both fundamental 
-$RR_{\rm {ab}}$- and first overtone -$RR_{\rm {c}}$- pulsators. 
This identification was facilitated by the empirical evidence 
that the pulsation amplitude and shape of the light curve 
of the two groups are quite different. 
The same outcome does not apply to high amplitude $\delta$ Scuti stars
(McNamara 2000) and classical Cepheids, since fundamental (F) and overtone
pulsators do show quite similar light curves nor the pulsation amplitudes
can be safely adopted to identify the pulsation mode (Santolamazza et al.
2000). 

The occurrence of FO pulsators among classical Cepheids was originally 
suggested by Pel \& Lub (1978), but their existence was a matter of 
concern till few years ago (Bohm-Vitense 1994) and it was definitively 
settled by the large sample of variable stars collected by the 
microlensing experiments (MACHO, EROS, OGLE). These teams also 
reported the detection of some tens of double-mode Magellanic Cepheids 
pulsating simultaneously in the first and in the second overtone 
(Beaulieu et al. 1997; Alcock et al. 1999; Udalski et al. 1999a, U99a).  
Even though, the existence of pure SO Cepheids was originally 
predicted by Stobie (1969) only two candidates 
(Galaxy, Burki et al. 1986; LMC, Alcock et al. 1999) have been 
proposed. Useful hints on the selection criteria to single out SOs  
were suggested by Antonello \& Kanbur (1997, AK97) 
on the basis of nonlinear SO models. However, the existence 
and the observational properties of these objects were rooted to 
the ground by Udalski et al. (1999b, U99b) who found a 
dozen low-amplitude, short-period SOs in a sample of 2300 SMC Cepheids. 

The main aim of this investigation is to present Cepheid full amplitude, 
convective SO models and compare theoretical predictions with observed 
SO Cepheids in the SMC. 
We selected these variables because we are interested in testing 
whether nonlinear, convective models account for the pulsation behavior of 
relatively metal-poor Cepheids. In fact, it has been recently suggested 
that current turbulent convective (TC) models do not account for energy 
dissipation sources in metal-poor Cepheids (Feuchtinger et al. 2000). 
At the same time, we are also interested in assessing whether 
SOs can be safely adopted to estimate stellar distances. Finally, we also 
computed nonlinear FO models, to figure out how the instability 
strip changes when moving toward lower luminosities. 
  
\section{Theoretical models and data} 

The full amplitude, nonlinear convective models presented in this 
investigation were constructed by adopting the input physics and physical 
assumptions discussed in Bono, Caputo, \& Marconi (1998, BCM) 
and in Bono, Marconi \& Stellingwerf (1999, BMS). In contrast 
with BCM and BMS we assumed a vanishing efficiency of turbulent 
overshooting in the regions where the superadiabatic gradient attains 
negative values. In fact, nonlinear, convective RR Lyrae models 
constructed by adopting this assumption account for the luminosity 
variation over a full cycle of U Com a field $RR_c$ variable  
(Bono, Castellani, \& Marconi 2000). Numerical experiments performed 
by adopting the same input parameters adopted by BMS suggest 
that the stability and the shape of the light curve of  F Cepheid 
models are marginally affected by this assumption. However, the 
luminosity amplitudes of both first and second overtones 
roughly decrease by 24\%. 

To cover the observed period range of SO Cepheids we constructed 
four different sequences of models at fixed chemical composition 
Y=0.25, Z=0.004 and stellar masses equal to 3.25, 3.5, 3.8, and 
4.0 $M_{\odot}$. 
The luminosity of these models was fixed according to the 
mass-luminosity (ML) relation adopted by BCM and BMS. Recent  
calculations (Bono et al. 2000) based on up-to-date canonical 
evolutionary models which cover a wide range of He and metal 
contents support this choice. To assess how the topology of the 
instability strip changes when moving toward lower luminosities, 
the previous models 
were implemented with three series of FOs with stellar masses equal 
to 5.0, 5.5, and 5.8 $M_{\odot}$. The static models were forced out 
of equilibrium by imposing a velocity amplitude of $5 km s^{-1}$ to 
the radial eigenfunctions. The nonlinear equations are integrated 
in time until the pulsation amplitudes show, over consecutive 
cycles, a periodic similarity of the order of $10^{-4}$.  

Table 1 summarizes the input parameters and selected nonlinear 
observables for the sequences of first and second overtone models.  
From left to right the first three columns give the stellar mass,
the luminosity and the effective temperature (K), while columns (4) 
to (6) list the period (d), the mean radius, and the fractional 
radius. Columns (7) to (9) give the bolometric (mag), the radial 
velocity (Km/s), and the effective temperature (K) amplitudes, 
while the last three columns list the I band amplitudes, 
as well as the Fourier parameters $\phi_{21}$ and $R_{21}$.     

\begin{table*}
\caption{Input parameters and selected nonlinear second and first 
overtone observables (Y=0.25, Z=0.004).}  
\begin{tabular}{cccccccccccc}
\hline
M$^1$ & $\log{L}^1$ & $T_e$  & Period & $R^1$ & $\Delta{R}/R$ & $\Delta{M_{bol}}$ & $\Delta{u}$ & $\Delta{T_e}$ & $A_I$ & $\Phi_{21}$ & $R_{21}$\\
 & &  K & d & & & mag & Km/s & K & mag & & \\
\hline
\multicolumn{12}{c}{Second Overtone}\\
3.25 & 2.49 & 6850 & 0.5037 &12.496 & 0.020  &0.362 &25.60 &600 &0.248 & 4.173 & 0.224 \\
3.25 & 2.49 & 6800 & 0.5177 &12.724 & 0.029 &0.500 &37.34 &850 &0.342 & 4.201 & 0.298 \\
3.25 & 2.49 & 6700 & 0.5406 &13.155 & 0.035 &0.516 &44.21 &950 &0.400 & 4.246 & 0.337 \\
3.25 & 2.49 & 6600 & 0.5664 &13.536 & 0.039 &0.482 &51.34 &750 &0.345 & 4.358 & 0.302 \\
3.25 & 2.49 & 6500 &0.5942 &13.904 &0.039 &0.385 &49.98 &600 &0.288 &4.633 & 0.248 \\
3.50 & 2.61 & 6750 & 0.6319 &14.869 & 0.019 & 0.307 &22.78 &500 &0.215 & 4.249 & 0.196 \\
3.50 & 2.61 & 6700 & 0.6437 &15.107 & 0.029 &0.424 &33.32 &700 &0.291 & 4.283 & 0.252 \\
3.50 & 2.61 & 6600 & 0.6747 &15.495 & 0.038 &0.480 &44.15 &750 &0.333 & 4.352 & 0.278 \\
3.50 & 2.61 & 6500 & 0.7067 &16.013 & 0.039 &0.409 &44.93 &600 &0.298 & 4.632 & 0.216 \\
3.80 & 2.74 & 6700 & 0.7826 &17.415 & 0.001 &0.015 &1.08& 50 & 0.011 & 4.592 & 0.105 \\
3.80 & 2.74 & 6600 & 0.8105 &18.004 &0.023 &0.333 &26.25& 550 & 0.247 & 4.438 & 0.173\\
4.00 & 2.82 & 6600 & 0.9140 &19.678 &0.005 &0.074 &5.34& 100 & 0.052 & 4.789 & 0.028 \\
\multicolumn{12}{c}{First Overtone}\\
3.25 & 2.49 & 6600 & 0.7049 & 13.573 & 0.077 & 1.099 & 83.56 & 1850 & 0.731 & 4.001  & 0.379  \\
3.25 & 2.49 & 6500 & 0.7377 & 14.041 & 0.082 & 0.975 & 86.65 & 1600 & 0.670 & 4.037 & 0.344 \\
3.25 & 2.49 & 6100 & 0.9015 & 15.811 & 0.066 & 0.365 & 57.16 & 700 & 0.268 & 4.765 & 0.281 \\
3.50 & 2.61 & 6600 & 0.8381 &15.599 & 0.075 & 1.114 & 79.09 &1900 &0.721 & 3.971 & 0.420 \\
3.50 & 2.61 & 6500 & 0.8848 &16.060 & 0.081 & 1.026 & 81.78 &1850 &0.656 & 3.972 & 0.410 \\
3.50 & 2.61 & 6400 & 0.9285 &16.656 & 0.084 & 0.873 & 81.18 &1850 &0.594 & 3.979 & 0.389 \\
3.50 & 2.61 & 6200 & 1.0256 &17.673 & 0.081 & 0.536 & 71.90 &1650 &0.397 & 4.225 & 0.320 \\
3.50 & 2.61 & 6100 & 1.0805 &18.167 & 0.074 & 0.440 & 62.65 &1550 &0.324 & 4.657 & 0.336 \\
3.50 & 2.61 & 6000 & 1.1331 &18.705 & 0.060 & 0.338 & 49.06 &600 &0.248 & 4.870 & 0.401 \\
3.80 & 2.74 & 6600 &1.0150 &18.105 &0.117  &0.956 &56.30 &1600 &0.635  &4.004  & 0.423 \\
3.80 & 2.74 & 6400 &1.1179 &19.202 &0.071  &0.857 &68.65 &1150 &0.673  &4.102  & 0.487 \\
3.80 & 2.74 & 6300 &1.1775 &19.918 &0.151  &0.798 &70.22 &1500 &0.568  &4.211  &0.474  \\
3.80 & 2.74 & 6200 &1.2375 &20.545 &0.069  &0.549 &62.81 &850 &0.445  &4.434  & 0.418 \\
3.80 & 2.74 & 6000 & 1.3720 &21.765 & 0.060 &0.440 & 47.96& 600 & 0.276 & 4.989 & 0.408 \\
4.00 & 2.82 & 6600 & 1.1403&19.787 & 0.023 &0.351 & 19.71& 550 & 0.248 & 4.016 & 0.230 \\
4.00 & 2.82 & 6000 & 1.5483&23.963 & 0.063 &0.392 & 50.81& 650 & 0.296 & 5.077 & 0.362 \\
4.00 & 2.82 & 5900 & 1.6221 &24.687 & 0.050 &0.291 & 38.10 & 500 & 0.224 & 5.058 & 0.434 \\
5.00 & 3.07 & 6400 & 1.7495&28.033 & 0.059 &0.830 & 47.61& 1300 & 0.583 &4.236
&0.437  \\
5.00 & 3.07 & 6200 & 1.9425&29.977 & 0.067 &0.811 & 64.62& 1200 & 0.601 & 4.444
& 0.456 \\
5.00 & 3.07 & 5900 & 2.2800 &32.969 & 0.062 &0.374 & 49.08 &600  & 0.296 & 5.259
 & 0.338 \\
5.50 & 3.32 & 6400 & 2.6553&37.230 & 0.002 &0.022 & 1.15& 50 & 0.016 & 5.036 & 0.445 \\
5.50 & 3.32 & 6200 & 2.9179&39.858 & 0.049 &0.553 & 36.03& 850 & 0.411 & 6.013 & 0.094 \\
5.50 & 3.32 & 6000 & 3.2626 &42.494 & 0.060 &0.512 & 45.58 &750  & 0.402 & 4.588 & 0.310 \\
5.80 & 3.40 & 6400 & 2.9627&40.931 & 0.001 &0.014 & 0.86& 50 & 0.011 & 5.124 & 0.208 \\
5.80 & 3.40 & 6200 & 3.2911&43.599 & 0.011 &0.132 & 7.44& 200 & 0.098 & 5.471& 0.235 \\
5.80 & 3.40 & 6000 & 3.6780 &46.756 & 0.060 &0.505 & 41.53 &700  & 0.401 & 4.739 & 0.077 \\
\end{tabular}

\medskip \noindent 
$^1$ Stellar masses, luminosities and radii are in solar units. 
\end{table*}

Fig. 1 shows the run of bolometric light and radial velocity curves 
over two consecutive cycles. Data plotted in this figure show two 
interesting features: i) both luminosity and velocity curves do show, 
in agreement with empirical evidence, smooth changes over the full 
pulsation cycle. However, few light curves show a small bump along the 
rising branch. It is not clear, whether this is a drawback of theoretical 
models, since one of the criteria adopted to select pure SO pulsators is 
a sinusoidal luminosity variation along the pulsation cycle. Note that 
the 9 pure SO pulsators we selected (see below) according to the Wesenheit 
index present light curves that are somewhat noisy, and therefore the bump 
can be barely detected. ii) SO luminosity, radius, and velocity 
amplitudes are approximately a factor of two smaller than FO amplitudes. 
On the other hand, the temperature amplitudes are roughly a factor of 
three smaller (see column (9) in Table 1). This suggests that the SO 
luminosity changes are mainly governed by radius rather than by 
temperature variations. 
Note that current SO bolometric and radial velocity amplitudes appear 
to be smaller than predicted by AK97. However, AK97 adopted a slightly 
different He and metal contents (Y=0.29, Z=0.01 against Y=0.25, Z=0.004).   
 
\begin{figure}
\vbox to 90mm{ 
\caption {Bolometric light curves (left) and radial velocity
curves (right) along a full pulsation cycle versus phase. 
The curves refer to models constructed by adopting different masses 
(see labeled values). The nonlinear periods (d), and the effective 
temperatures (K) are listed in the left and in the right panel 
respectively.}
\vfil}
\label{fig1}
\end{figure}

We first consider the modal stability of the selected Cepheid models. 
In order to compare theory and observations the predicted instability 
edges were transformed 
into the observational plane by adopting the bolometric corrections 
and color-temperature relations derived by Castelli, Gratton, \& 
Kurucz (1997). 
Fig. 2 shows the topology of the Cepheid instability strip for the first
three radial modes in the $M_V, V-I$ Color Magnitude (CM) diagram. 

\begin{figure}
\vbox to 80mm{ 
\caption {Topology of the Cepheid instability strip for the first three 
radial modes in the CM diagram. Solid and dashed lines refer to blue and 
red edges respectively. The arrows mark the OR regions between first 
and second overtone (FO/SO) as well as between fundamental and first 
overtone (F/FO).}
\vfil}
\label{fig2}
\end{figure}

A glance at the data plotted in this figure shows that the region
within which the SO is unstable widens when moving from high 
to low luminosities, and indeed the temperature width increases 
from 100 K to almost 500 K. The same outcome applies to the OR region, 
i.e. the region in which both SO and FO pulsators are unstable, and 
indeed its temperature width for M=3.25 $M_\odot$ is roughly equal 
to 200 K. Moreover, predicted period ratios -$P_{SO}/P_{FO}$- in this 
region range from 0.799 to 0.805 and are in fair agreement with empirical
values ($0.802\div0.809$, U99a). The widening of the SO instability strip 
is mainly caused by the fact that the decrease in luminosity causes a 
substantial decrease in the SO total kinetic energy when compared with 
the FO. In fact, along the FOBE the ratio between SO and FO kinetic 
energy ranges from 0.66 at  $M/M_\odot$=5 to 0.55 at $M/M_\odot$=4, and to 
0.39 at $M/M_\odot$=3.25. A decrease in the SO total kinetic energy implies 
a decrease in the pulsation inertia of the envelope, and in turn in the 
amount of energy necessary to destabilize this mode (Gilliland et al. 1998).  

Data plotted in Fig. 2 also show that the width of the FO instability
region remains almost constant when moving from 5 to 3.25 $M_\odot$,
whereas the width of the F region decreases from 400 K to 200 K.
The narrowing of the F region is mainly due to a sharp change in the
slope of the red edge at lower luminosities. This finding supports the
empirical evidence recently brought out by Bauer et al. (1999) that when
moving toward lower luminosities the number of F pulsators decreases and
the slope of the PL relation for this mode changes. A more quantitative
analysis of this effect will be addressed in a forthcoming paper
(Bono et al. 2000). We also note that, the width of the OR region between
F and FO remains constant and for $M \le 3.8 M_\odot$ it becomes a factor
of two smaller than the OR region between FO and SO. This difference
supplies a qualitative explanation to the empirical evidence that among
double-mode SMC Cepheids the 25\% are F/FO, while the other are FO/SO
variables (U99a). We also find that, the predicted period ratios $P_{FO}/P_F$
in this region range from 0.725 to 0.746 that are once again in very good
agreement with the observed ones ($0.728\div0.746$, U99a). Moreover,
it is worth underlining that the total width of the instability strip, in
the above mass range, remains constant and roughly equal to 1000 K. This
means that when accounting for F, FO, and SO variables the edges of the
Cepheid instability strip in the SMC are approximately parallel. 
It goes without saying that the occurrence of two OR regions makes fainter 
SMC Cepheids key objects to constrain predictions on the topology of the 
instability strip and on modal selection (Kollath 1999).

To properly identify pure SO pulsators in SMC, U99b adopted 
three different selection criteria, namely the Fourier parameters 
together with the location in both the Color-Magnitude and the PL 
diagram. The range of plausible $R_{21}$ and $\phi_{21}$ values was 
selected, according to Alcock et al. (1999), on the basis of Fourier 
parameters of SO light curves in double-mode pulsators, while the 
period range was selected according to the $P_{SO}/P_{FO}$ period ratio.    
By using these criteria U99b found 13 bona fide pure SO pulsators. 
We can now check whether the selection criteria they adopted are 
supported by current nonlinear convective models. Fig. 3 shows 
the comparison between the Wesenheit index of pure SO pulsators 
detected by U99b and predicted SO models (solid line). 
Although few objects seem to be systematically brighter/cooler 
than predicted the agreement between theory and observations is 
quite satisfactory. 

\begin{figure}
\vbox to 80mm{ 
\caption{Wesenheit index vs period for pure  
SO pulsators (open circles, U99b), SO component in FO/SO Cepheids 
(open squares, U99a), pure FO pulsators (filled circles, U99b), 
FO component in FO/SO Cepheids (filled squares, U99a). Open triangles 
show the new selected SO pulsators, while solid and dashed lines 
display the predicted Wesenheit indices for second and first 
overtones respectively. Data were plotted by assuming DM=19.1 mag.}   
\vfil}
\label{fig3}
\end{figure}

Note that current predictions suggest that some FO variables which do not 
obey to the selection criteria adopted by U99b could be pure SO pulsators, 
since they are located in the same region covered by theoretical models. 
As a preliminary but plausible 
assumption we decided to include in the sample of pure SO pulsators the 
FO variables located within three $\sigma$ ($\sigma_W$=0.03 mag, Udalski 
2000, private communication) from the predicted SO loci. Interestingly,
enough we find that the color 
distribution of the new candidates (open triangles) are quite similar 
to the V-I colors of FO/SO Cepheids, and indeed they range from 
0.45 to 0.65 (U99a). 
Finally, we also included in Fig. 3 the FO/SO double-mode Cepheids 
to test whether current models do account for their distribution. 
The adopted periods, mean magnitudes and colors refer to data given 
by U99a (see their Appendix C). 
The physical assumptions adopted for constructing current nonlinear,
convective models are somehow supported by the agreement between theory 
and observations. However, before any firm conclusion on the TC model 
currently adopted can be reached predictions based on nonlinear convective 
models should account for the entire observational scenario, and in
particular for mixed-mode variables, and Fourier coefficients of F and 
FO Cepheids (Feuchtinger et al. 2000). 

Fig. 4 shows the comparison between predicted and observed luminosity  
amplitudes (top panel) as well as $\phi_{21}$ (bottom panel) values. 
Data plotted in this figure refer to the Fourier fit of the I band 
light curves of SO components in FO/SO Cepheids and to pure SO Cepheids. 
Predicted amplitudes (filled triangles) up to $\log P\approx-0.15$ are 
systematically larger than the observed ones, whereas at longer periods 
they are quite similar to empirical ones.  

\begin{figure}
\vbox to 80mm{ 
\caption{Comparison between predicted and observed amplitudes (top) 
and $\phi_{21}$ Fourier parameters (bottom). Amplitudes and 
$\phi_{21}$ values refer to the I band light curves. Theoretical and 
empirical curves were fitted with Fourier series which include 8 terms. 
The symbols are the same as in Fig. 3.} 
\vfil} 
\label{fig4}
\end{figure}

Theoretical $\phi_{21}$ values are, within current uncertainties, in 
fair agreement with observed data for pure SO variables. The same 
outcome applies for the bulk of the $\phi_{21}$ values of the SO 
components in FO/SO Cepheids. The error bars for these objects 
are larger because the SO is the secondary mode and therefore 
their luminosity amplitudes are relatively small.
Predicted and empirical $R_{21}$ parameters are also in plausible 
agreement but unfortunately the latter ones are affected by large 
errors. 
Data plotted in the top panel support the evidence that the luminosity  
amplitudes can be safely adopted for selecting pure SO pulsators, since 
in this period range they are systematically smaller than FO amplitudes
(Fig. 5). At the same time, the selection criterium based on the 
Fourier parameters of SO components in FO/SO Cepheids does not seem 
appropriate, since the SOs cover the same region covered by FOs 
(Figs. 4 and 5). 

Fig. 5 shows the same data of Fig. 4 but they refer to pure FOs and to 
FO components in FO/SO Cepheids. Once again predicted amplitudes are, 
at short periods, systematically larger than the observed ones, whereas 
for periods longer than $\log P\approx0$ the theoretical amplitudes attain 
values closer to the empirical ones. 
On the other hand, predicted $\phi_{21}$ values agree quite well with
observations for periods shorter than $\log P=0.4$. Note that theoretical
models for $M/M_\odot$=5.5 and 5.8 show for $0.4 < \log P \le 0.5$ a
well-defined minimum in the pulsation amplitude, whereas predicted
$\phi_{21}$ values are, in the same period range, larger than the
empirical ones.
More detailed calculations are necessary to assess whether this feature 
is connected with the occurrence of a Hertzsprung progression among FOs
(Kienzle et al. 1999). 
  
\begin{figure}
\vbox to 80mm{
\caption{Same as Fig. 4 but for FO pulsators.}  
\vfil}
\label{fig5}
\end{figure}


To disclose whether SOs are good standard candles we derived on the basis 
of current SO models the following analytical Period-Luminosity-Color (PLC) 
relation: 

\begin{tabular}{llll} 
$M_V=$& 3.961 &$-3.905\log{P}$ &$+3.250(V-I)\;\;\;$ \\ 
        & $\pm 0.005$  & $\pm 0.019$        & $\pm 0.054$  \\
\end{tabular}

where $M_V$ and (V-I) are weighted mean intensities transformed into 
magnitudes, $P$ is the period (d), and the standard deviation is 
$0.004$. The spread of this relation is small because the temperature
width of SO instability strip is substantially smaller than F and   
FO strips.

The top panel of Fig. 6 shows the projection of the PLC relation 
into a plane. Interestingly enough, we find that by fitting the sample 
of pure SO pulsators together with the FO/SO Cepheids the SMC 
distance modulus is $19.11\pm0.08$ mag. The uncertainty accounts for 
errors on both photometry and reddening  but does not account for depth 
effects. Owing to the lack of individual reddening estimates we adopted, 
according to Caldwell \& Coulson (1986), a mean reddening value 
of $E(B-V)=0.054\pm0.029$. The current SMC distance 
estimate supports the distance determination derived by Bono et al. 
(1999, $DM=19.19\pm0.17$ mag) on the basis of the fundamental PLC (V, B-V) 
relation and data available in the literature. Our distance estimate 
is in very good agreement with the SMC distance derived by 
Laney \& Stobie (1994) on the basis of empirical PL relations in 
four different photometric bands i.e. $DM=18.94\pm0.04$ mag 
(internal error), by Kovacs (2000) using double-mode SMC Cepheids 
($DM=19.05\pm0.13$ mag), and by Groenewegen (2000) using both the Wesenheit 
index ($DM=19.08\pm0.11$ mag) and the $PL_K$ relation ($DM=19.04\pm0.17$ mag).    

\begin{figure}
\vbox to 80mm{
\caption{Period-Luminosity-Color (V,V-I) relations for second (top) and 
first (bottom) overtone pulsators. Solid lines show theoretical relations, 
while the symbols are the same as in Fig. 3.}
\vfil}
\label{fig6}
\end{figure}


Forced by the above result we decided to estimate the SMC distance 
by adopting the theoretical PLC relation for FOs. By taking into 
account current and old (Bono et al. 1999) models we find the 
following PLC (V, V-I) relation: 

\noindent 
$M_V = 3.61(\pm 0.03)\, -3.85(0.02) \log P \, +3.33(\pm 0.09)(V-I)$ 
%
where the symbols have their usual meaning and the standard deviation 
is $0.03$. The fit of this relation to both pure FOs and  FO/SO 
Cepheids (see bottom panel of Fig. 6) supplies $DM=19.16\pm0.19$ mag. 
The mean magnitudes and colors of these variables were corrected by 
adopting the same mean reddening adopted for SO variables. 
Even though FOs cover different period and luminosity ranges, the new 
distance agrees, within the errors, with the distance based on SOs. 
At the same time, both of them are at odds with SMC distances 
based on red clump stars ($DM=18.63\pm0.07$ mag) and on field RR Lyrae 
stars ($DM=18.66\pm0.16$ mag) derived by Udalski (1998) and by 
Udalski et al. (1998).  
Finally, we note that FO models for $M/M_\odot$=5 allow us to test the 
dependence
of predicted FO edges on the adopted TC calibration. Interestingly enough,
the new edges differ by less than 100 K when compared with the edges
predicted by Bono et al. (1999) on the basis of the old TC calibration
(Bono \& Stellingwerf 1994). Therefore, it turns out that the new TC
calibration marginally affects predicted PL and PLC relations. 

\section{Conclusions}

We have presented and discussed theoretical predictions concerning the 
pulsation properties of SO pulsators among SMC Cepheids. The comparison 
between theory and observations brought out the following results:   
i) current metal-poor, nonlinear, convective models account for 
the pulsation behavior of pure first and second overtone Cepheids. 
In fact, predicted $\phi_{21}$ parameters are in fair agreement with 
empirical data (U99a,U99b).  
ii) Theoretical predictions show that the OR region, i.e. the region in 
which both first and second overtones are unstable, widens in temperature 
when moving from higher to lower luminosities. This evidence supplies        
a qualitative explanation to the large number of FO/SO double-mode 
Cepheids detected by OGLE in the SMC.  
iii) Predicted period ratios -$P_{SO}/P_{FO}$, $P_{FO}/P_F$- of the 
models located in the two OR regions are in good agreement with 
observed values for FO/SO and F/FO double-mode SMC Cepheids (U99a).  
   
Interestingly enough we find that both pure SO variables and the SO 
components in FO/SO Cepheids are good distance indicators, and indeed 
the fit of the theoretical PLC (V,V-I) relation to empirical SMC data 
supplies a distance modulus of $19.11\pm0.08$ mag. This result is 
further strengthened by the fact that the distance modulus we obtain 
by adopting the predicted PLC relation for FO variables is 
$19.16\pm0.19$ mag. These distance estimates are based on a mean 
reddening correction, since the SMC Cepheid sample collected by 
OGLE lack of individual reddening measurements. Current distance 
estimates are, within the uncertainties, at odds with the so-called 
short distance scale.    


\section*{Acknowledgments}
We thank V. Castellani for a critical reading of an early draft 
of this manuscript and D. Laney for useful discussions on 
reddening estimates. We also acknowledge A. Udalski for many helpful 
advice to retrieve data from the OGLE database, M. Groenewegen for 
communicating his results in advance of publication, and an anonymous 
referee for useful comments that improved the content of this paper.
This work was supported by MURST -Cofin98- under the project: 
"Stellar Evolution".


\label{lastpage}


\begin{thebibliography}{99}
\bibitem{} Alcock C. et al., 1999, AJ, 117, 920
\bibitem{} Antonello E., Kanbur S. M., 1997, MNRAS, 286, 33 (AK97)
\bibitem{} Bauer F., et al. 1999, A\&A, 348, 175 
\bibitem{} Beaulieu J. P. et al., 1997, A\&A,  318, L47
\bibitem{} Bohm-Vitense E., 1994, AJ, 107, 673  
\bibitem{} Bono G., Caputo F., Cassisi S., Marconi M., 
Piersanti L., Tornamb\`e A., 2000, ApJ, accepted, astro-ph/0006251 
\bibitem{} Bono G., Caputo F., Castellani V., Marconi M. 1999, ApJ, 512, 711
\bibitem{} Bono G., Caputo F., Marconi M., 1998, ApJ, 497, L43 (BCM)
\bibitem{} Bono G., Castellani V., Marconi M., 2000, ApJ, 532, L129
\bibitem{} Bono G., Marconi M., Stellingwerf R. F. 1999, ApJS, 122, 167 (BMS)
\bibitem{} Bono G., Stellingwerf R. F. 1994, ApJS, 93, 233  
\bibitem{} Burki G. et al., 1986, A\&A, 168, 139
\bibitem{} Caldwell J. A. R., Coulson I. M. 1986, MNRAS, 218, 223
\bibitem{} Castelli F., Gratton R. G., Kurucz R. L. 1997, A\&A, 324, 432
\bibitem{} Eddington A. S., 1926 in The Internal Constitution of the Stars, 
Cambridge University Press, Cambridge, p. 203  
\bibitem{} Feuchtinger M. U., Buchler J. R., Kollath Z., 2000, ApJ, submitted,
astro-ph/0005230
\bibitem{} Gilliland R. L., Bono G., Edmonds P. D., Caputo F., Cassisi S.,
Petro L. D., Saha A., Shara M. M., 1998, ApJ, 507, 818  
\bibitem{} Groenewegen M. A. T. 2000, A\&A, accepted, astro-ph/0010298
\bibitem{} Kienzle F., Moskalik P., Bersier D., Pont F. 1999, A\&A, 341,818 
\bibitem{} Kollath Z. 2000, in IAU Coll. 176, The Impact of Large-Scale 
Surveys on Pulsating Star Research, eds. Szabados, L. Kurtz, D., ASP, 
San Francisco, p. 356 
\bibitem{} Kovacs G. 2000, A\&A, 360, L1  
\bibitem{} Laney C. D., \& Stobie R. S. 1994, MNRAS, 266, 441 
\bibitem{} Ledoux P., \& Walraven Th. 1958, in Handbuch der Physik, 51, 353
\bibitem{} McNamara D. H. 2000, PASP, 112, 1096 
\bibitem{} Pel J. W., Lub J., 1978 in The HR diagram - The 100th anniversary 
of H. N. Russell, Reidel, Dordrecht, p. 229
\bibitem{} Santolamazza P. Marconi M., Bono G., Caputo F., Cassisi S.,
\& Gilliland R. L. 2000, ApJ, submitted  
\bibitem{} Schwarzschild M., 1941, ApJ, 94, 245
\bibitem{} Stobie R. S., 1969, MNRAS, 144, 511
\bibitem{} Udalski A., 1998, AcA, 48, 113 
\bibitem{} Udalski A., Pietrzynski G.; Wozniak P., Szymanski M.,
Kubiak, M., Zebrun , K 1998, ApJ, 509, L25  
\bibitem{} Udalski A. et al., 1999a, AcA, 49, 1 (U99a)  
\bibitem{} Udalski A. et al., 1999b, AcA, 49, 45 (U99b) 
\end{thebibliography}
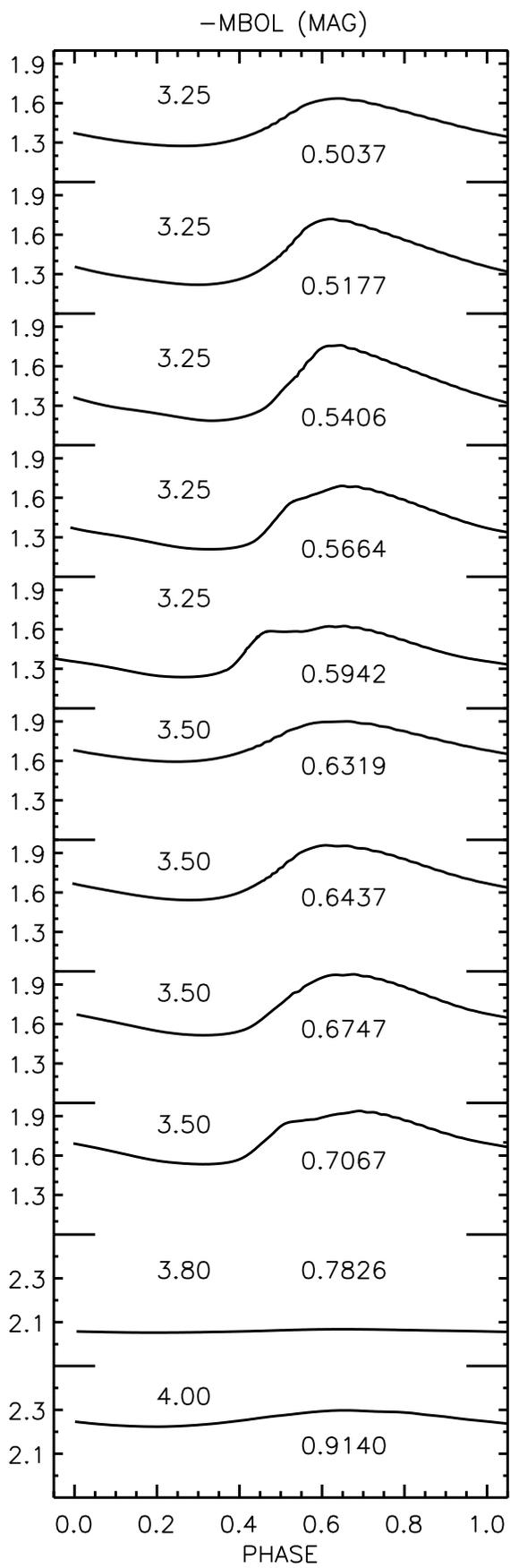
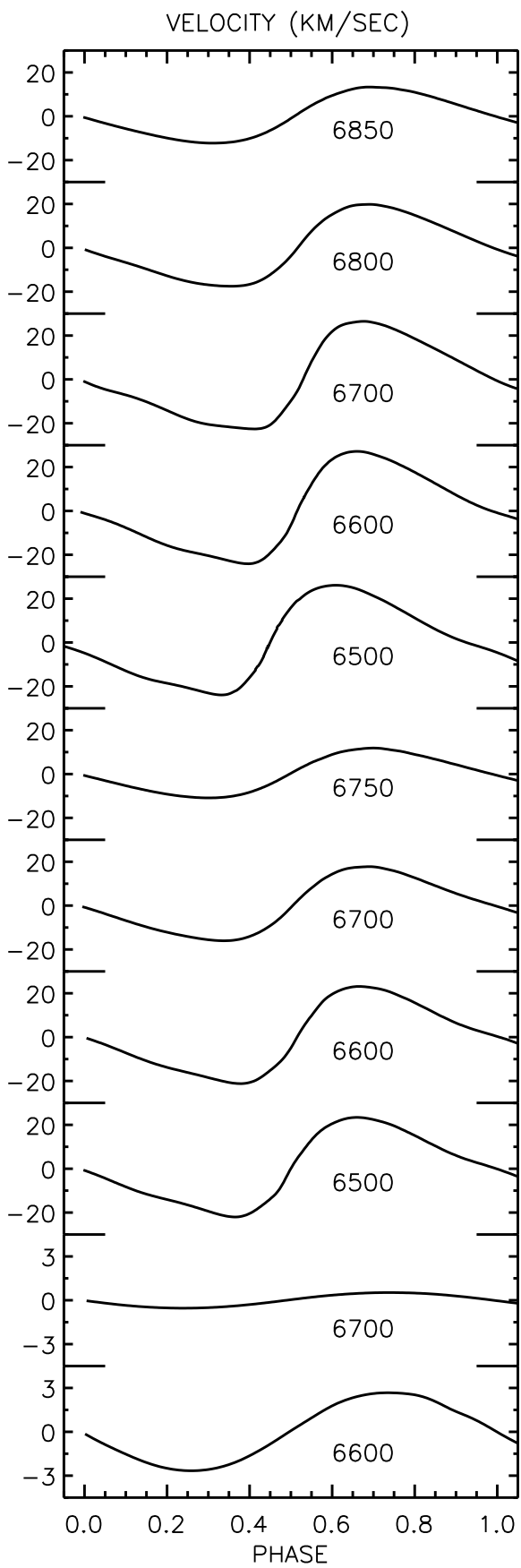
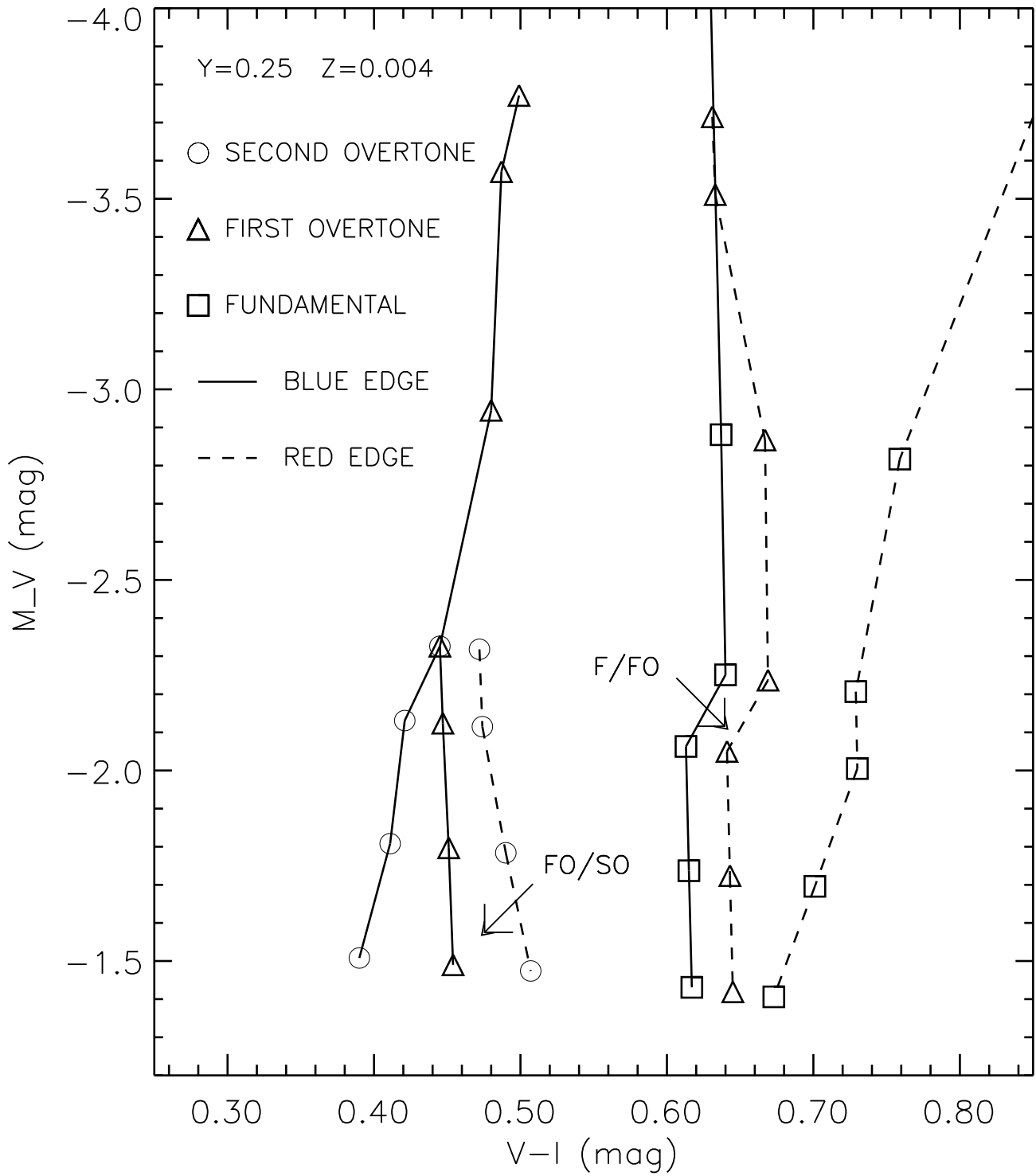
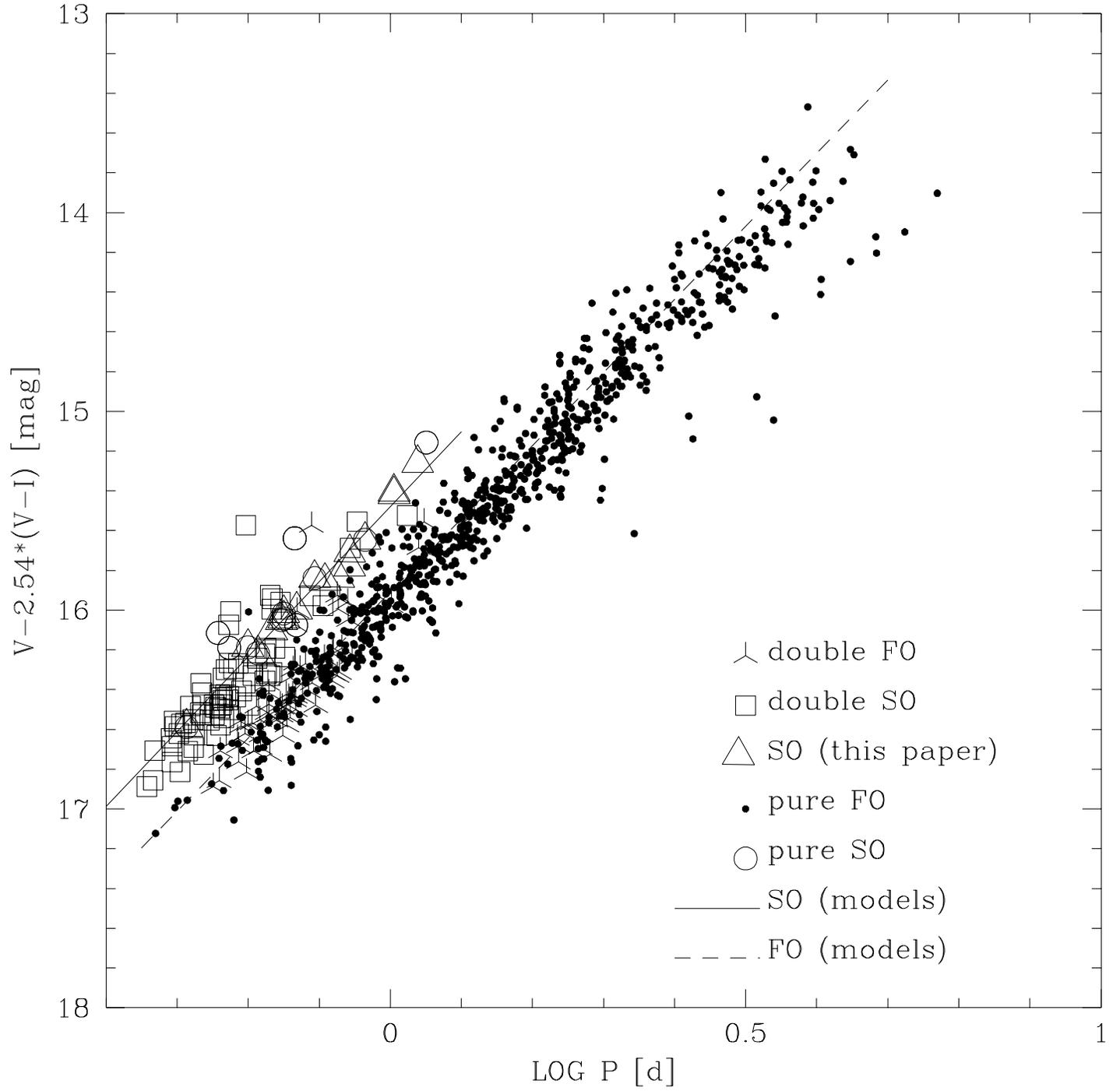
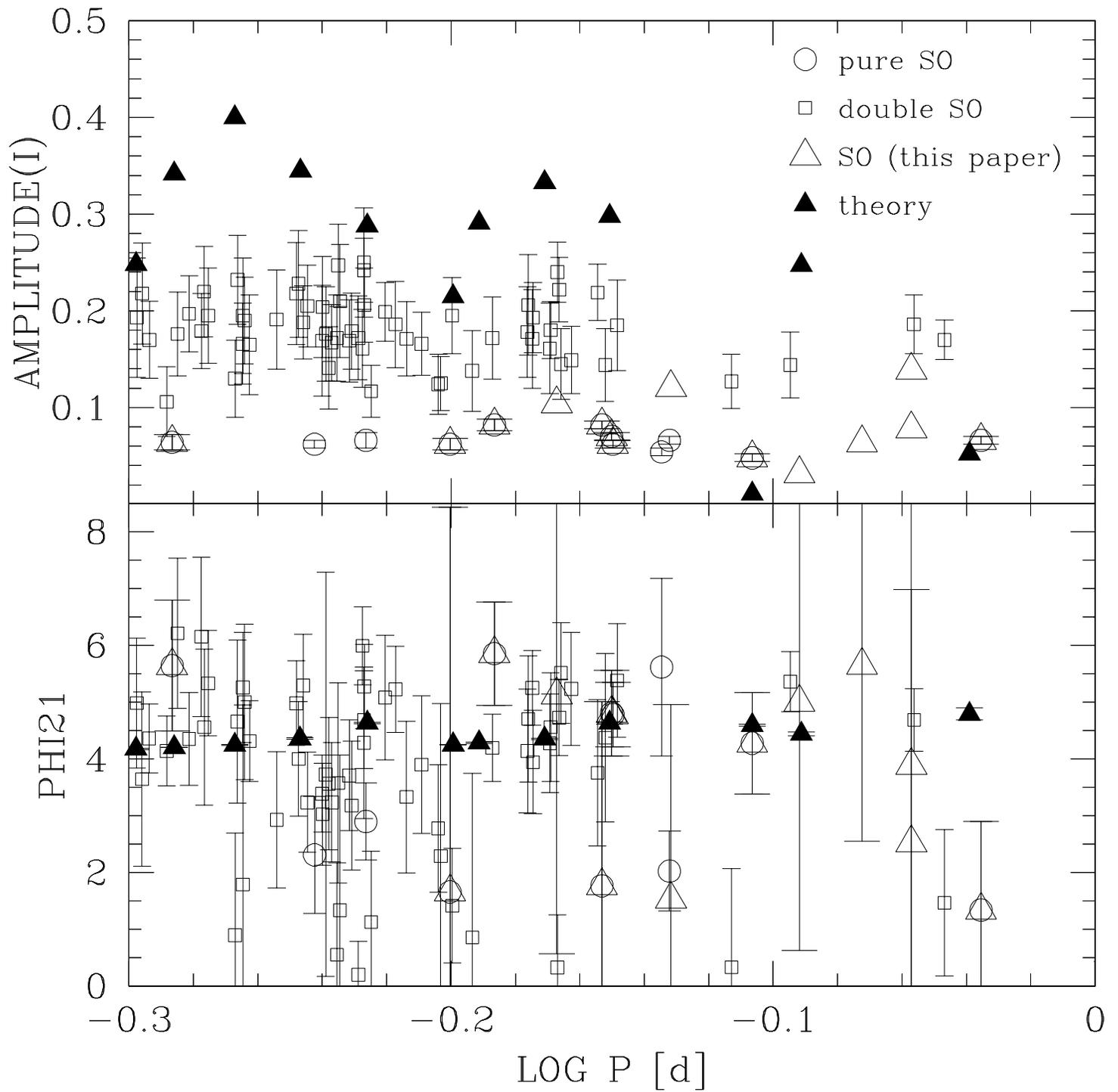
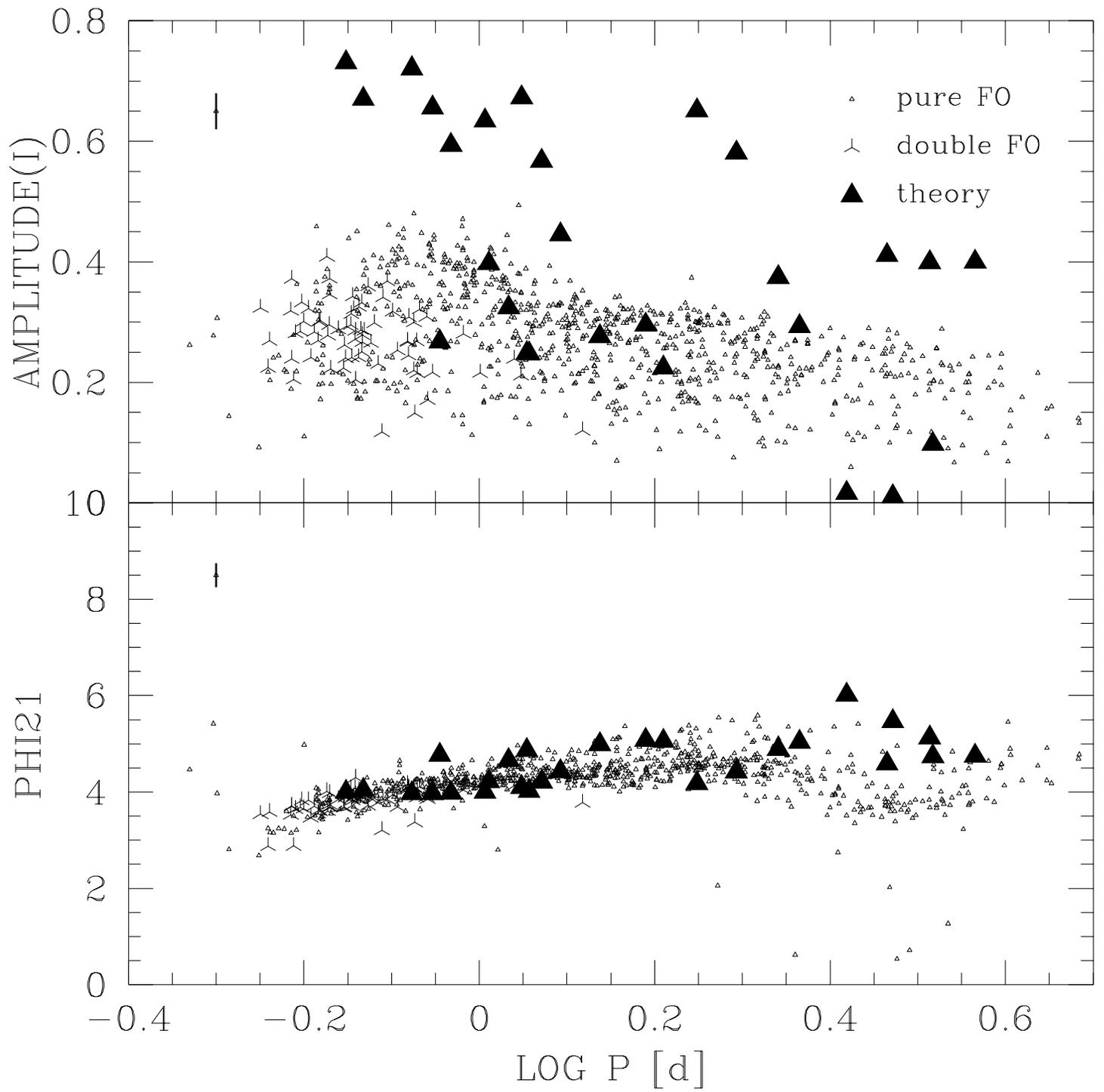
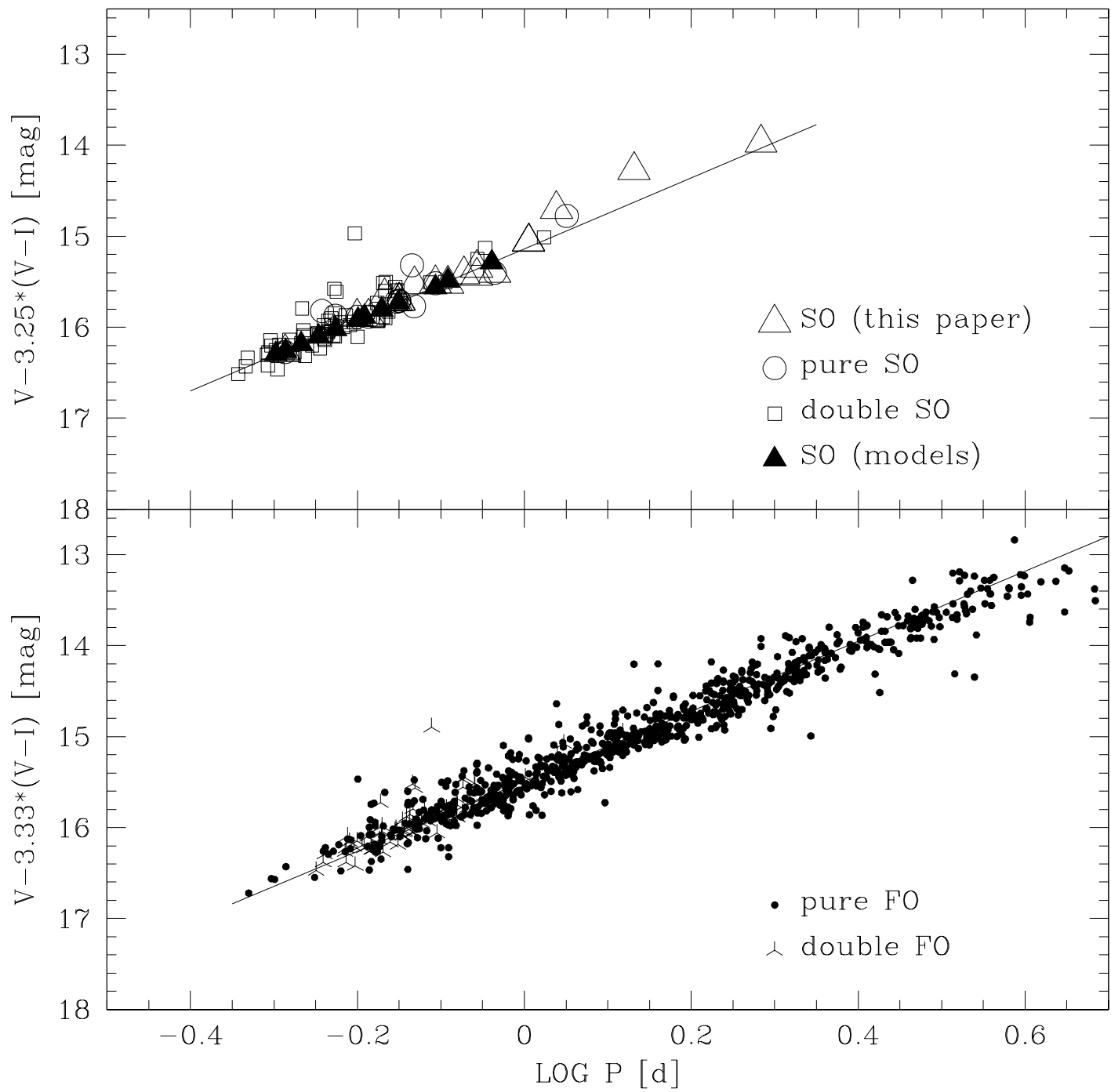
\end{document}